\documentclass[useAMS,usenatbib,referee]{biom}
\usepackage[]{lineno}
\usepackage{algorithm}
\usepackage{color}
\usepackage{amsmath} 
\usepackage{standalone}
\usepackage{amsfonts}
\usepackage{pdflscape}
\usepackage{graphicx}
\usepackage{subfig}
\usepackage[normalem]{ulem}
\usepackage{soul}
\usepackage{url}
\usepackage{multirow}
\usepackage{arydshln}
\graphicspath{{./fig/}}
\newcommand{\bs}[1]{\boldsymbol{#1}}
\newcommand{\PM}{\mbox{PM$_{2.5}$}}

\title{Multivariate spectral downscaling for $\PM$ species}

\author{Yawen Guan$^{*}$\email{yguan12@unl.edu}\\
	Department of Statistics, University of Nebraska\\
	\textbf{ Brian J Reich}\\
	Department of Statistics, North Carolina State University\\
	\textbf{James A Mulholland}\\
	School of Civil \& Environmental Engineering, Georgia Tech\\
	\textbf{Howard H Chang}\\
	Department of Biostatistics and Bioinformatics, Emory University
}

\begin{document}
	\begin{abstract}
		Fine particulate matter ($\PM$) is a mixture of air pollutants that has adverse effects on human health. Understanding the health effects of $\PM$ mixture and its individual species has been a research priority over the past two decades. However, the limited availability of speciated $\PM$ measurements continues to be a major challenge in exposure assessment for conducting large-scale population-based epidemiology studies. The $\PM$ species have complex spatial-temporal and cross dependence structures that should be accounted for in estimating the spatiotemporal distribution of each component. Two major sources of air quality data are commonly used for deriving exposure estimates: point-level monitoring data and gridded numerical computer model simulation, such as the Community Multiscale Air Quality (CMAQ) model. We propose a statistical method to combine these two data sources for estimating speciated $\PM$ concentration. Our method models the complex relationships between monitoring measurements and the numerical model output at different spatial resolutions, and we model the spatial dependence and cross dependence among $\PM$ species. We apply the method to combine CMAQ model output with major $\PM$ species measurements in the contiguous United States in 2011.
	\end{abstract}
	
	\maketitle

\section{Introduction}
Fine particulate matter ($\PM$) is a mixture of air pollutants with aerodynamic diameters 2.5 micrometers and smaller. Exposure to $\PM$ concentration has been linked to a variety of health problems, such as respiratory and cardiovascular diseases \citep{Atkinson660}, as well as adverse birth outcomes \citep{LI2017596}. However, while the health effects of the total $\PM$ mass are well understood, there are limited and contradictory findings in the associations of different PM chemical components with the same health endpoints, potentially due to the differences in accuracy of exposure information \citep{Grahame2009}. The main challenges in predicting species exposure arise from limited observations and their complex correlation. The emission sources of $\PM$, such as motor vehicles, coal-fired power plants and other manufacturing facilities are widespread and spatially heterogeneous. That, combined with the complex chemical reactions among pollutants result in highly temporally and spatially varying composition of $\PM$. There is a need in developing computationally efficient statistical methods that can capture the complex structure and make reliable maps of species concentrations, which will facilitate epidemiologic research on identifying more reliable associations between specific species and health endpoints.

A common approach to produce accurate maps is to fuse disparate data sources such as deterministic computer model output and observations from monitoring stations \citep[see, e.g.,][for reviews]{Gotway2002, CressieWikle2011}. For single pollutant predictions, numerous spatiotemporal methods have been developed to downscale gridded model output to point-level \citep[e.g.,][]{Fuentes2005,Berrocal2010,McMillan2010}. Extensions of these methods have also been proposed to handle ensemble forecasts \citep[e.g.,][] {Berrocal2007,Feldmann2015,Schefzik2017}, distribution matching \citep[e.g.,][]{Clark2004,gel2004,Huang2019complete}, extreme value analysis \citep[e.g.,][]{mannshardt2010downscaling,reich2013extreme,bechler2015spatial}, and to account for forecast errors \citep{Berrocal2012}.  Less attention is paid for joint downscaling of multiple pollutants, where capturing the complex interactions and dependence among species are essential to improve accuracy. A few most relevant methods for our purposes are summarized here. \cite{berrocal2010bivariate} propose a bivariate spatial downscaler for $\PM$ and ozone that both accounts for bias in the model output for each type of air pollution and also borrows strength across pollutants to improve precision. \cite{choi2009multivariate}, \cite{crooks2014simultaneous}, and \cite{rundel2015data} jointly model speciated $\PM$ and total $\PM$ combining data from different monitoring networks. \cite{huang2018multivariate} jointly model several pollutants and their health effects. These methods correct for the additive and multiplicative biases in the numerical model, but pay little attention to such biases in different spatial scales of the numerical model. 

We propose a multivariate spectral downscaler for speciated $\PM$. We model the associations between the numerical model output and observations at multiple resolutions using a spectral analysis. This provides insights on the performance of the numerical model at different spatial scales and can potentially improve predictions by using only the appropriate scales. We further exploit the cross and spatial dependence among observations by joint modeling multiple pollutants to improve prediction. This is particularly useful for $\PM$ constituents, because their observations are more sparse than the total $\PM$ due to the limited number of monitors and lower sampling frequencies. The proposed method is fit using parallel MCMC \citep{ConsensusMCMC}, where inference based on the full data is made by combining posterior distributions based on small batches of the data set. This work is an extension of the univariate spectral downscaler proposed by \cite{Reich2014}, where their focus is prediction for a single pollutant. 

The rest of this paper is organized as follows. Section 2 introduces the data sources used in the proposed downscaler. Section 3 provides details on the spectral analysis, the univariate and multivariate spectral downscalers and model fitting. Section 4 presents the data analysis results for the contiguous US in 2011. We conclude with Section 5 giving a brief summary and some final remarks on the proposed approach. Details on exploratory data analysis to verify model assumptions and a simulation study are presented in the Supplementary Materials.

\section{Data}
We consider daily 24-hour average estimates and measurements of total $\PM$ and its major constituents: elemental carbon (EC), organic carbon (OC), nitrate (NO$_3$), sulfate (SO$_4$) and ammonium (NH$_4$). Two sources of information were obtained from the United States Environmental Protection Agency (EPA): monitoring data for the Air Quality System and model output from the Community Multiscale Air Quality (CMAQ) model. The raw station data are right skewed, therefore, we analyze the log transformed station data and apply the same transformation on the CMAQ data (see Supplementary Materials S.1 for plots of the station data before and after transformation). 

\begin{figure}
	\centering
	\subfloat[Station $\PM$ (log) on Jan 1st, 2011\label{fig:stnpm2.5}]{\includegraphics[width=8cm,height=4cm,trim={0cm 0cm 0cm 0cm},clip]{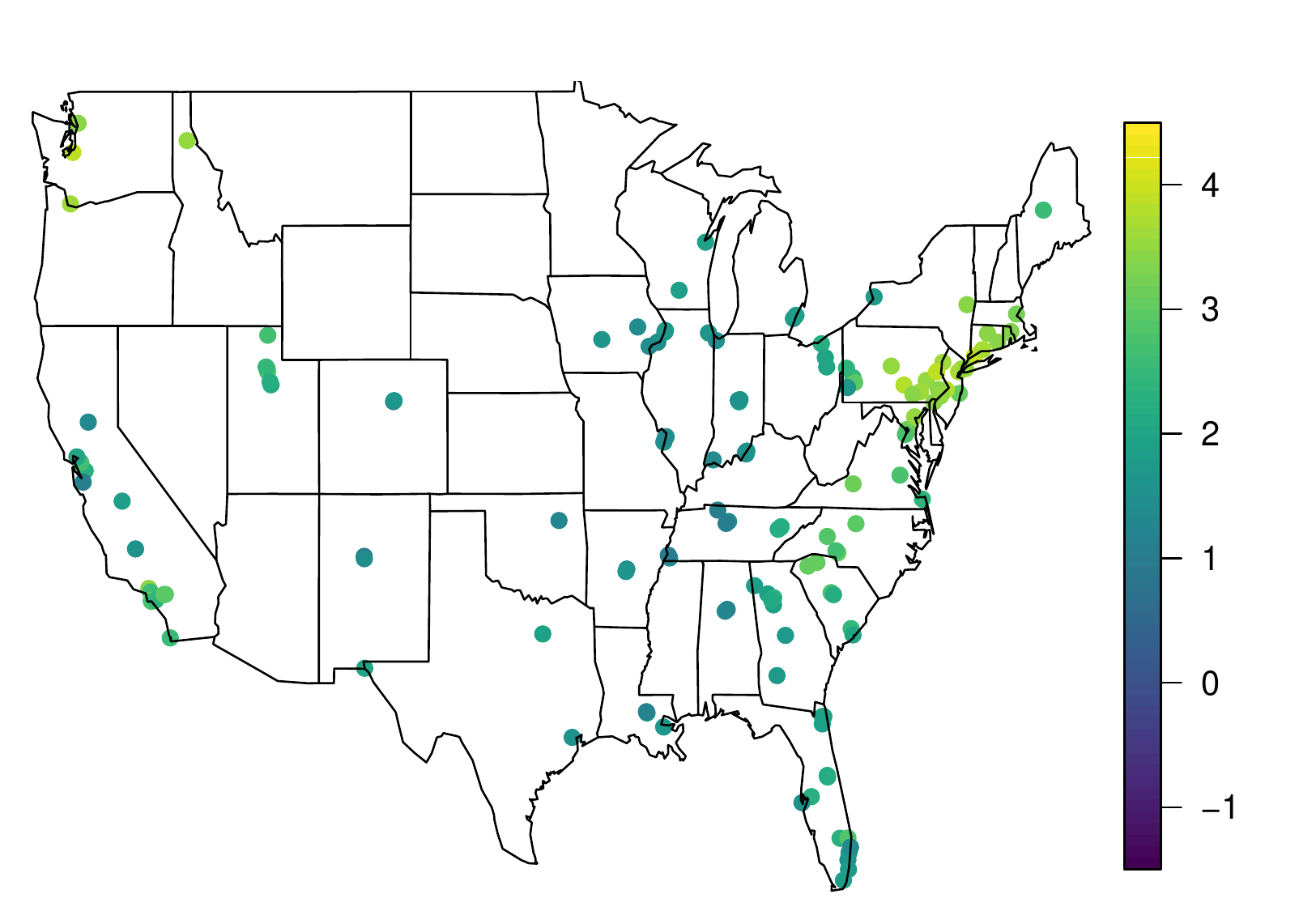}}
	\subfloat[CMAQ $\PM$ (log) on Jan 1st, 2011\label{fig:cmaqpm2.5}]{\includegraphics[width=8cm,height=4cm,trim={0cm 0cm 0cm 1cm},clip]{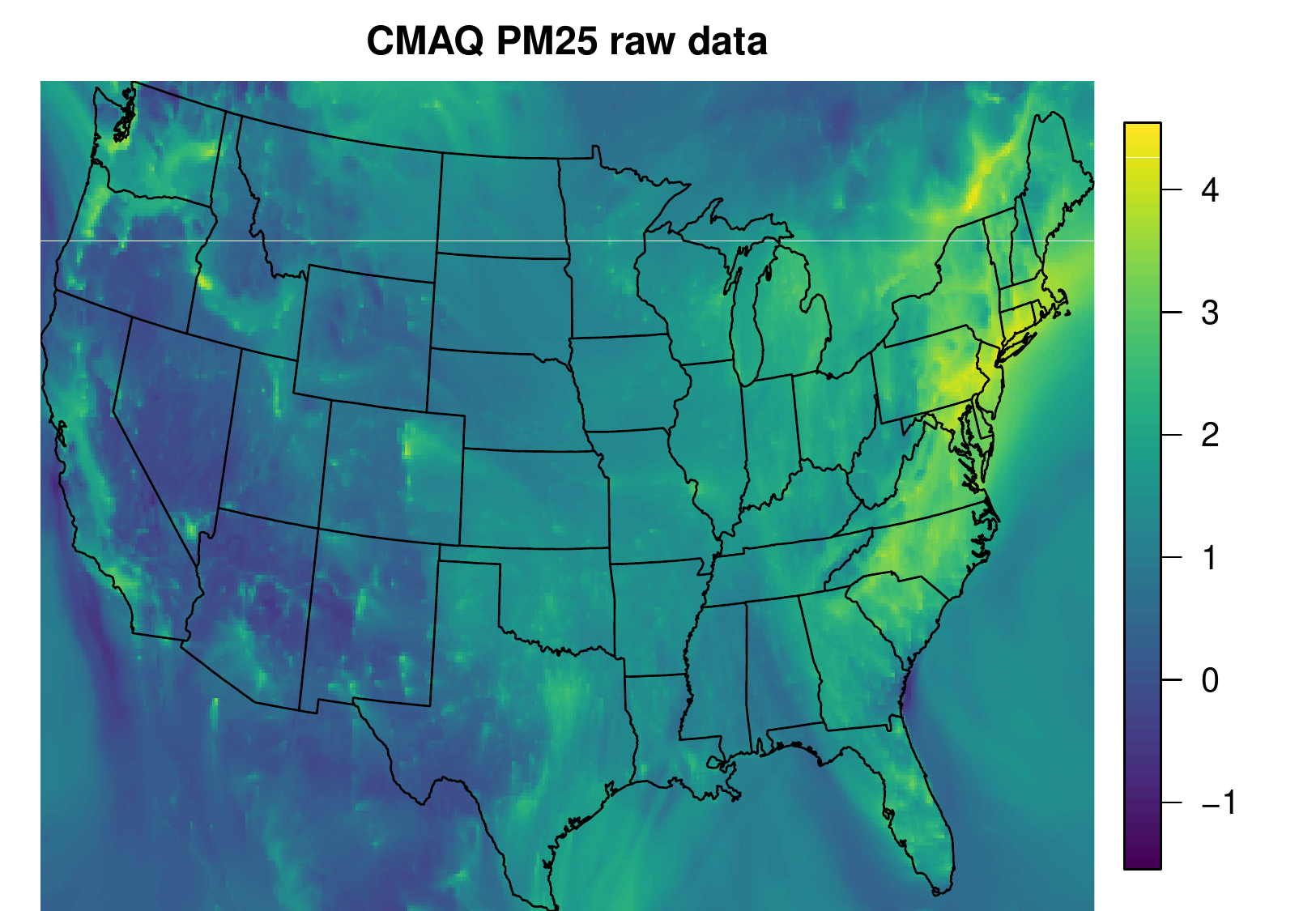}}
	
	\subfloat[Station EC (log) on Jan 1st, 2011\label{fig:stnec}]{\includegraphics[width=8cm,height=4cm,trim={0cm 0cm 0cm 0cm},clip]{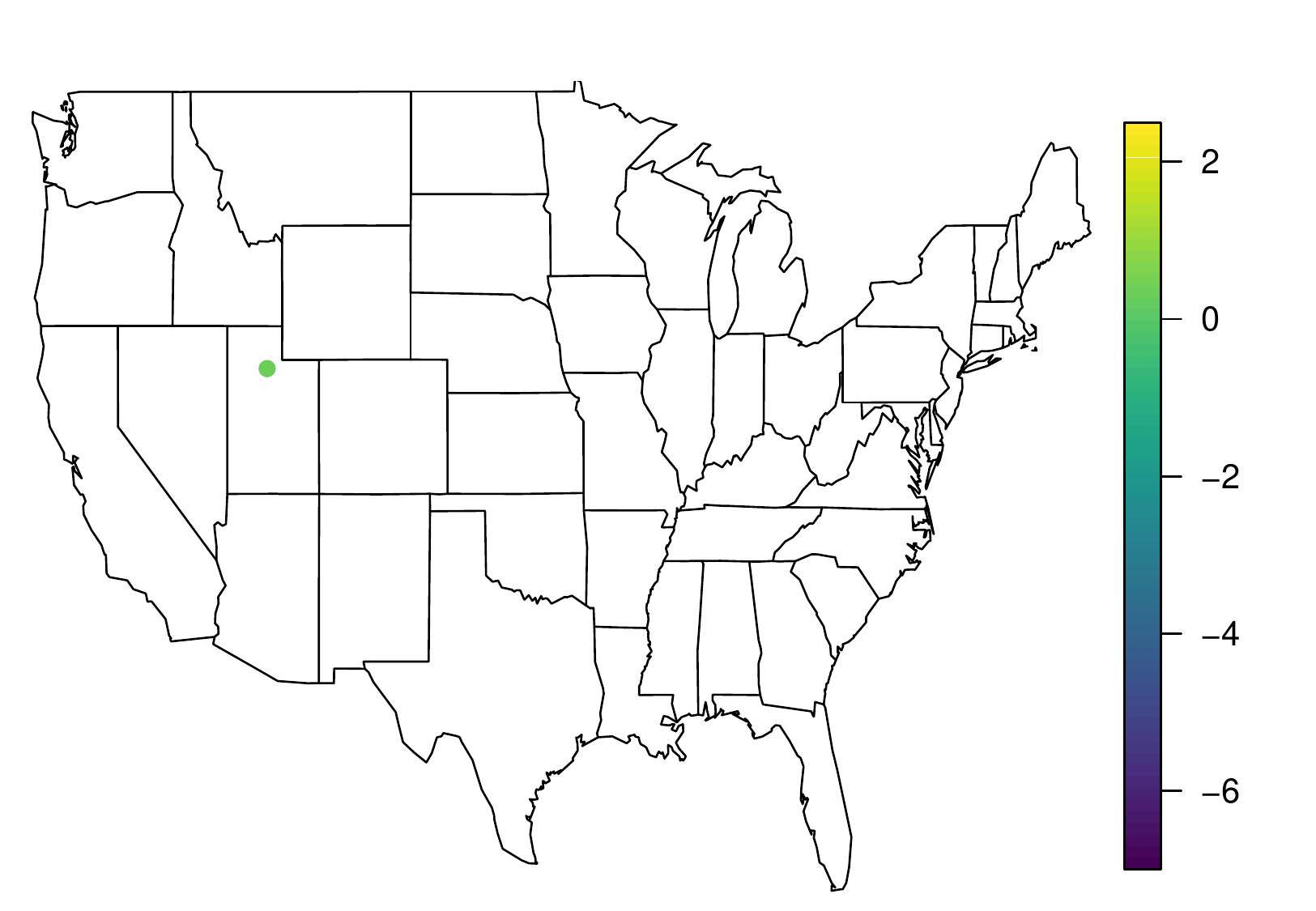}}
	\subfloat[CMAQ EC (log) on Jan 1st, 2011\label{fig:cmaqec}]{\includegraphics[width=8cm,height=4cm,trim={0cm 0cm 0cm 1cm},clip]{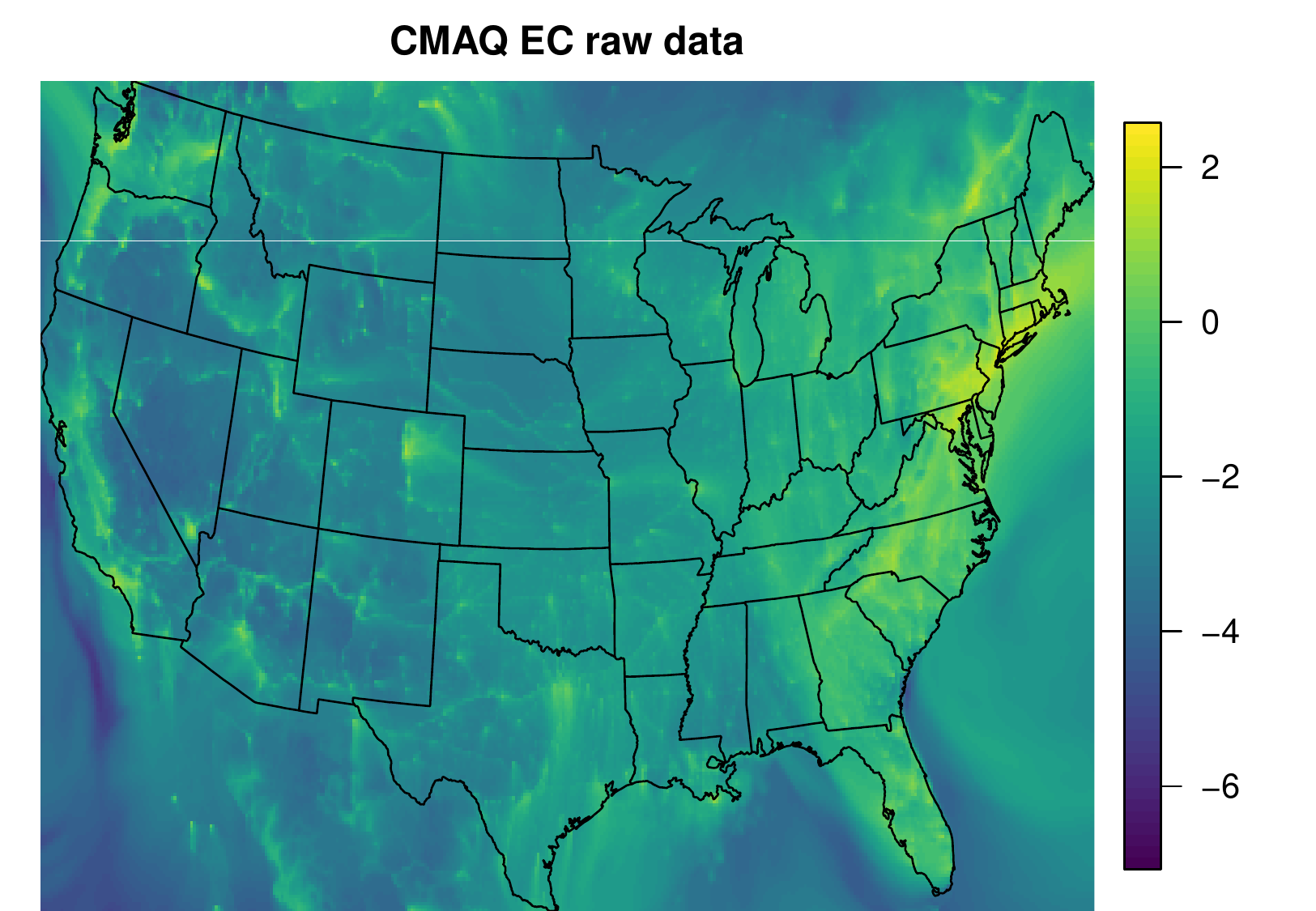}}
	
	\subfloat[Station $\PM$ (log) on Jan 3rd, 2011\label{fig:stnpm2.5_3}]{\includegraphics[width=8cm,height=4cm,trim={0cm 0cm 0cm 0cm},clip]{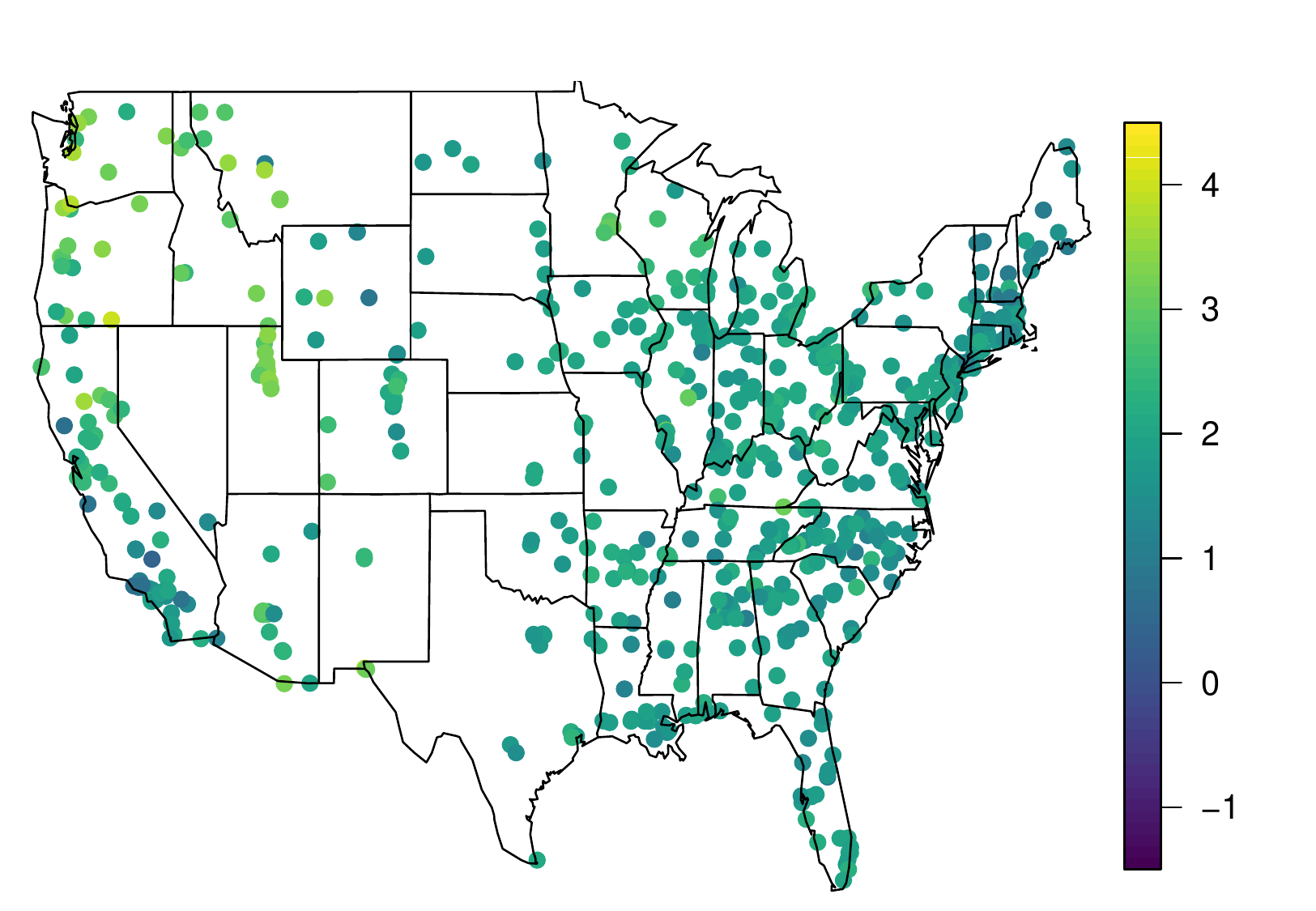}}
	\subfloat[CMAQ $\PM$ (log) on Jan 3rd, 2011\label{fig:cmaqpm2.5_3}]{\includegraphics[width=8cm,height=4cm,trim={0cm 0cm 0cm 1cm},clip]{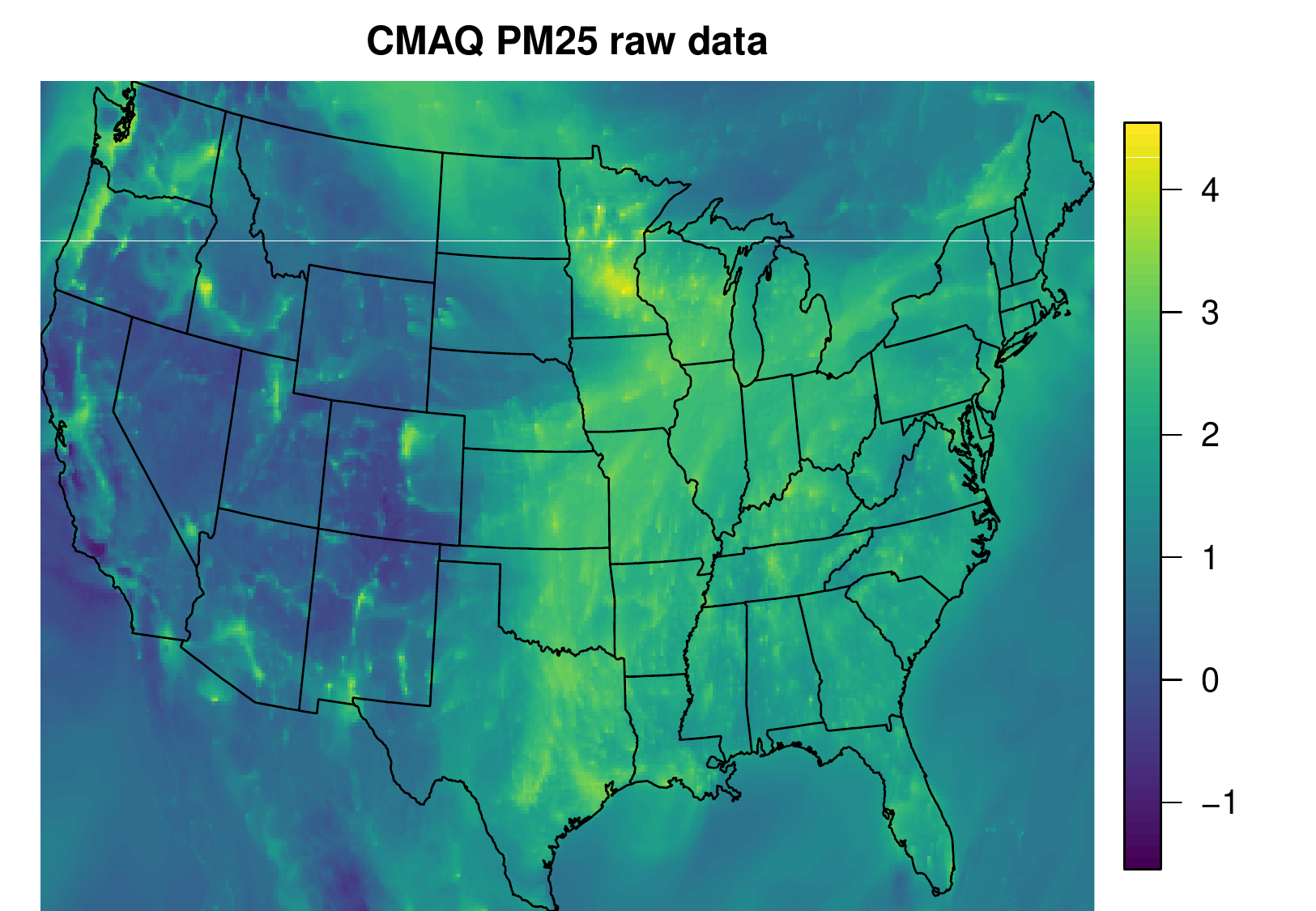}}
	
	\subfloat[Station EC (log) on Jan 3rd, 2011\label{fig:stnec_3}]{\includegraphics[width=8cm,height=4cm,trim={0cm 0cm 0cm 0cm},clip]{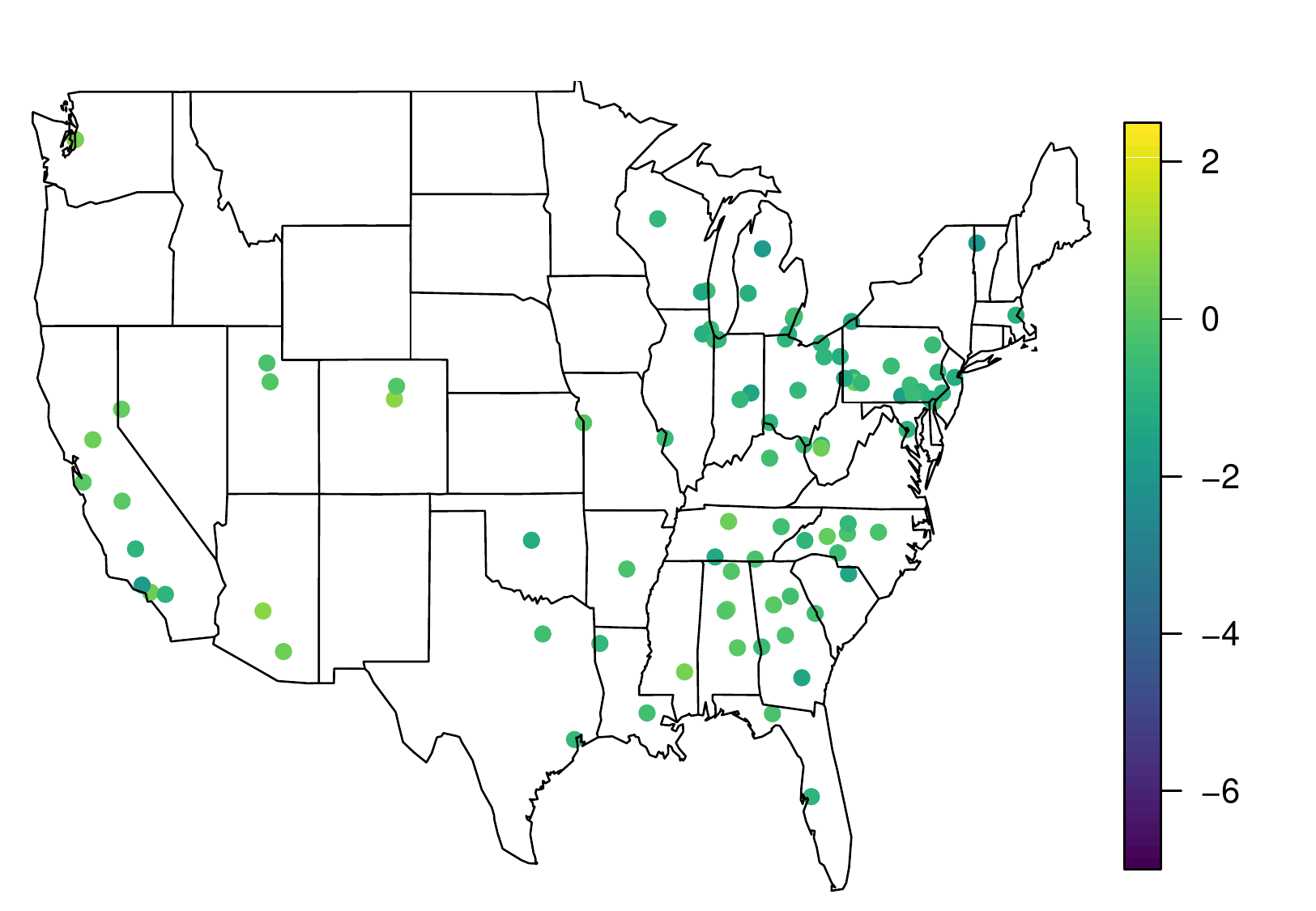}}
	\subfloat[CMAQ EC (log) on Jan 3rd, 2011\label{fig:cmaqec_3}]{\includegraphics[width=8cm,height=4cm,trim={0cm 0cm 0cm 1cm},clip]{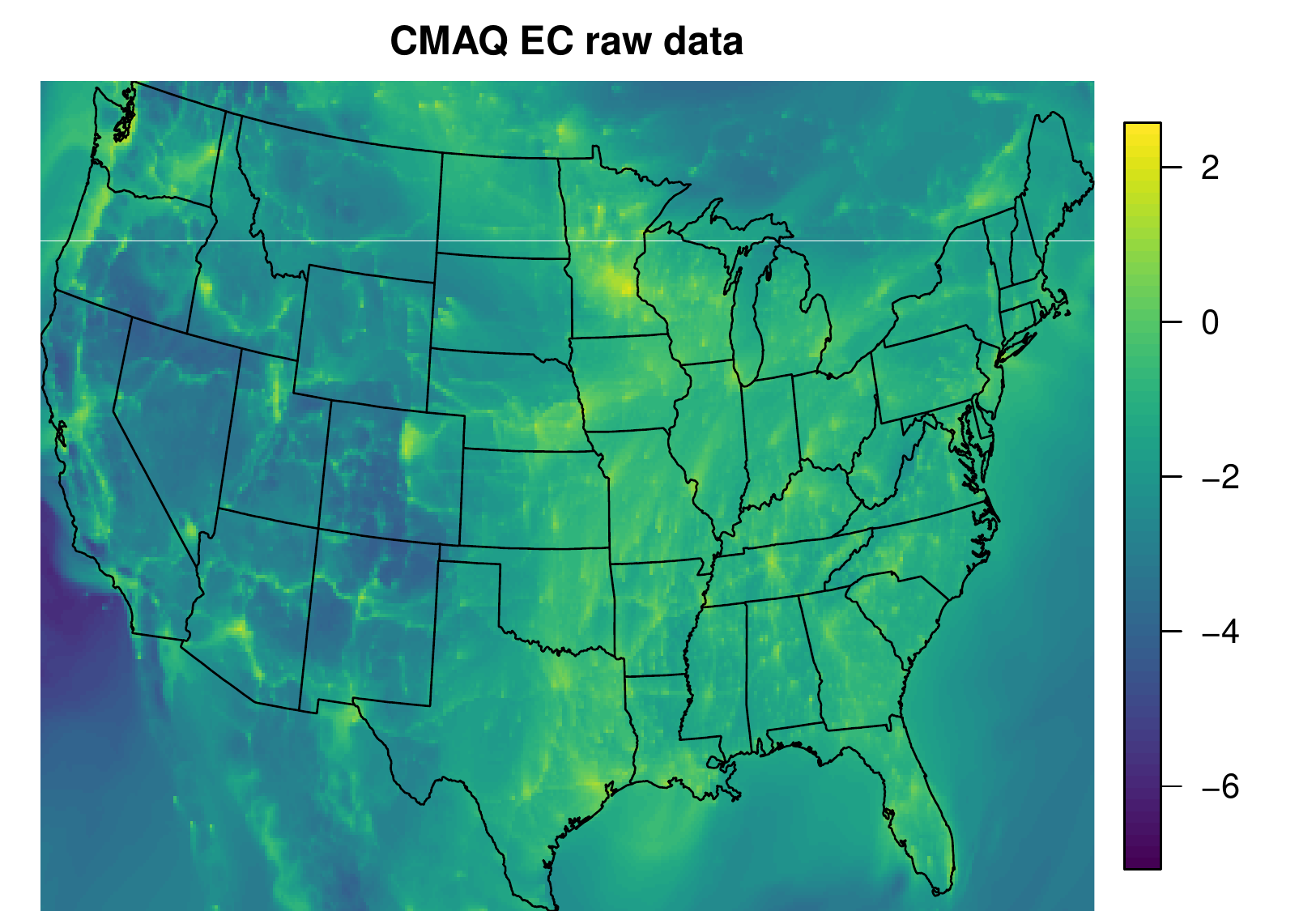}}
	
	\caption{Monitoring data of $\PM$ and EC (left), and CMAQ output of $\PM$ and EC concentrations (right) in unit log($\mu g/m^3$), for Jan 1 and 3, 2011.}
	\label{fig:cmaq_stn}
\end{figure}

In our analysis, monitoring data come from 845 stations (including all PM2.5 monitors for regulatory purposes and monitors from SPECIATE network) located throughout the contiguous United States (a map of station locations is shown in the Supplementary Material S.1). Not all stations measure the same pollutants, among them, 734 measure only total $\PM$, 15 measure only the constituents, 95 measure both total $\PM$ and its constituents, and 1 measures only  NO$_3 $ and  SO$_4 $. For total $\PM$, these measurements are taken daily, 1-in-3 or 1-in-6 days; for speciated $\PM$, these measurements are mostly taken 1-in-3 or 1-in-6 days. Figure \ref{fig:cmaq_stn} (left) shows the monitoring data for both $\PM$ and EC on Jan 1st and Jan 3rd, 2011. The $\PM$ total and its species measurements are spatially sparse and not all are measured at the same locations (referred to as spatial misalignment), and the sampling locations of $\PM$ species vary greatly across days.

CMAQ is a deterministic model that simulates daily pollutant concentrations accounting for emission sources, and complex atmospheric chemistry and physics. It provides gridded maps of multipollutant concentrations, which can serve as proxy data for locations without a monitoring station. The version v5.02 used in this paper provides output at 12 km spatial resolution, on a grid of 299 by 459 cells for the contiguous US. As an example, Figure \ref{fig:cmaq_stn} (right) shows the CMAQ output for both $\PM$ and EC on Jan 1st and Jan 3rd, 2011.

\section{Modeling Framework}
Here we present a multivariate spectral downscaler that models the complex relationship among pollutants and two data sources. We begin by reviewing the univariate spectral downscaler proposed by \cite{Reich2014}, where they analyzed the correlation between CMAQ and monitoring data of ozone in the spectral domain. For simplicity, we first present purely spatial multivariate downscaling and then discuss the temporal aspect.
	
\subsection{Spectral Representation}\label{s:spectralbasic}
We provide a brief background on spectral analysis; a more comprehensive review is given by \cite{handbook:spectral}. A stationary spatial process $X(\bs{s})$ has a spectral representation,
\begin{equation}\label{eqn:integral}
	X(\bs{s}) =\int \exp(i\bs{\omega}^T\bs{s})Z(\bs{\omega})d\bs{\omega},
	\end{equation}
where $\bs{s} \in R^2$ is a spatial location, $\bs{\omega} \in R^2$ is a frequency and $Z(\bs{\omega})$ is the spectral process corresponding to $ X(\bs{s}) $. If we further assume that $X(\bs{s})$ is a mean-zero Gaussian process, then  $Z(\bs{\omega})$ is also Gaussian with $E[Z(\bs{\omega})] = 0$, $V[Z(\bs{\omega})] = \sigma^2f(\bs{\omega})$, and $Cov[Z(\bs{\omega}),Z(\bs{\omega}')] = 0$. The smoothness of the spatial process is determined by the spectral density $f(\bs{\omega})$, which weights the frequencies. If more weight is given to small frequencies then the spatial process will be smoother; if more weight is given to large frequencies then the spatial process will be relatively rougher. Spectral methods are particularly useful for decorrelating spatial processes and decomposing spatial variation over different scales (or resolutions). 

To illustrate how the spectral representation decomposes the spatial signal into different scales, we extract and plot the signal corresponding to different $||\bs{\omega}||$. For spatial process observed on a grid $\mathcal{D} (m_1\times m_2)$ with $M=m_1m_2$ equally spaced locations, we can compute $ Z(\bs{\omega}_l), l = 1,\dots,M$ by taking inverse discrete Fourier transform of $X$. The available Fourier frequencies are $\bs{\omega}_l \in 2\pi\mathcal{J}_M$, for $\mathcal{J}_M = m_1^{-1}\{\lfloor -(m_1-1)/2\rfloor,\dots,m_1-\lfloor m_1/2 \rfloor \} \times m_2^{-1}\{\lfloor -(m_2-1)/2\rfloor,\dots,m_2-\lfloor m_2/2 \rfloor \}.$ Then the spatial signal with frequencies $||\bs{\omega}||\in[L,U)$ is approximated by the integral in \eqref{eqn:integral} integrated over [L,U)
\begin{equation}\label{eqn:filterX}
	\tilde{X}_{[L,U)}(\bs{s}) =\sum_{||\bs{\omega}_l|| \in [L,U)} \exp(i\bs{\omega}_l^T\bs{s})Z(\bs{\omega}_l),
	\end{equation} 
	
\begin{figure}
	\centering
	\subfloat[$ \tilde{X}_{[0,\pi/5)} $\label{fig:filterXa}]{\includegraphics[width=0.5\textwidth,page=2 ]{PM25filter10_264_zoom_fixed.pdf}}
	\subfloat[$ \tilde{X}_{[\pi/5,2\pi/5)} $\label{fig:filterXb}]{\includegraphics[width=0.5\textwidth,page=3 ]{PM25filter10_264_zoom_fixed.pdf}}
	
	\subfloat[$ \tilde{X}_{[4\pi/5,\pi)} $\label{fig:filterXc}]{\includegraphics[width=0.5\textwidth, page=6]{PM25filter10_264_zoom_fixed.pdf}}
	\subfloat[$ \tilde{X}_{[7\pi/5,8\pi/5)} $\label{fig:filterXd}]{\includegraphics[width=0.5\textwidth,page=9]{PM25filter10_264_zoom_fixed.pdf}}
	
	\subfloat[CMAQ $\PM$ (log)\label{fig:filterX0}]{\includegraphics[width=0.5\textwidth,page=1 ]{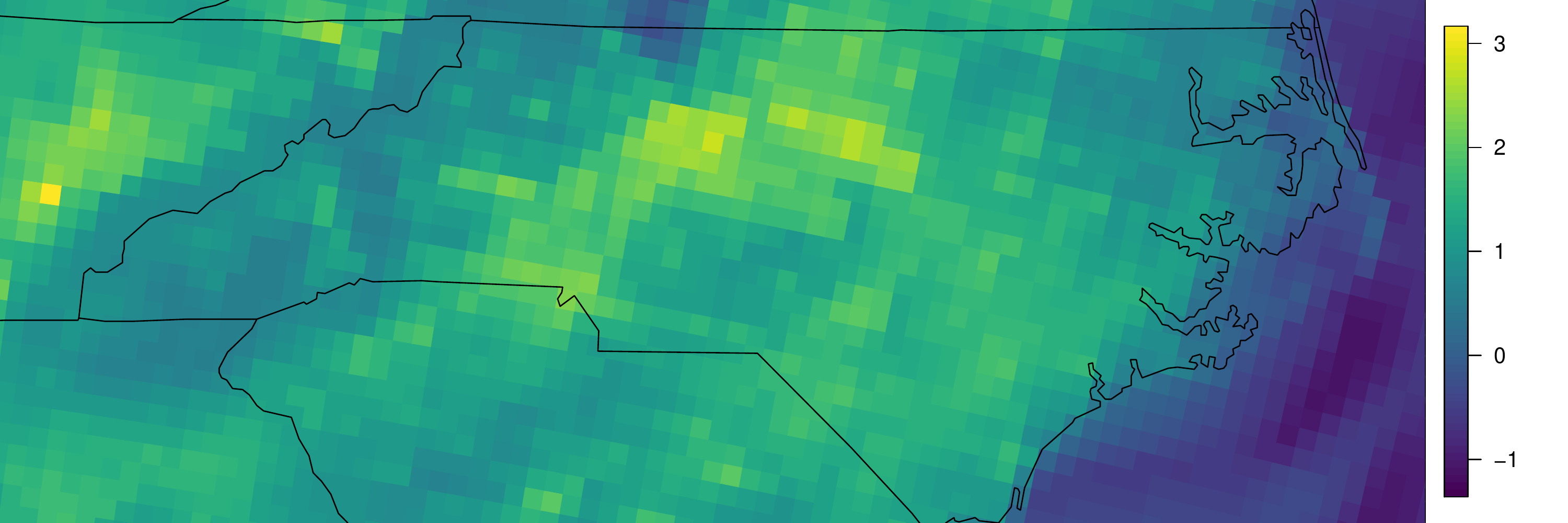}}
	
	\caption{Filtered CMAQ $\PM$ on Sept 21, 2011 by dividing the available frequency with magnitude of $ [0, \sqrt{2}\pi] $ into 8 equal-width bins with width $\pi/5$; {four of them are shown here}.  The signals at low frequencies $ [0,\pi/5) $ (b) resemble the large spatial-trend ($>$120km), and at $ [\pi/5,2\pi/5) $ (c) resemble spatial scale between 60 to 120km, while at higher frequencies $ [4\pi/5,\pi) $ (d) resemble small scale (24$\sim$30km) information. Finally, the signals corresponding to the high frequencies (d) resembles large-scale features, as a consequence of aliasing. For comparison, CMAQ $\PM$ is shown in (e).}
	\label{fig:filterX}
\end{figure}

For CMAQ gridded multipollutant concentrations, we explore the signals at different scales for each pollutant to see their contributions to the spatial process. An example of the extracted signals for CMAQ $\PM$ on Sept 21, 2011 is shown in Figure \ref{fig:filterX}; here, we divide the range of available frequency magnitude $ [0, \sqrt{2}\pi] $ into 8 equal-width bins with width $\pi/5$, and filter the $\PM$ concentration to retain the signals corresponding to each bin. We see that the signals at low frequencies $ [0,\pi/5) $ (Figure \ref{fig:filterXa}) resemble the large spatial-trend of CMAQ $\PM$ (Figure \ref{fig:filterX0}), while at higher frequencies (Figures \ref{fig:filterXb} - \ref{fig:filterXc}) resemble small-scale information (Figure \ref{fig:stnpm2.5_3}); finally, the signals corresponding to high frequencies (Figure \ref{fig:filterXd}) resembles large-scale features, as a consequence of the aliasing effect (see Supplementary Materials S.2). 
		
	\subsection{Univariate Spectral Downscaler}
	
	The univariate spectral downscaler of \cite{Reich2014} models the association between monitoring data and CMAQ for a single pollutant. Let $Y(\bs{s})$ be monitoring data and $ X(\bs{s}) $ be the CMAQ output at the grid cell containing the station location $\bs{s}$. A joint spectral model for two stationary mean-zero spatial processes is
	
	\begin{equation}
	\begin{aligned}
	X(\bs{s}) &= \int \exp(i\bs{\omega}^T\bs{s})Z_1(\bs{\omega})d\bs{\omega} \\
	Y(\bs{s}) &= \int \exp(i\bs{\omega}^T\bs{s})Z_2(\bs{\omega})d\bs{\omega} ,
	\end{aligned}
	\end{equation}
	where $Z_i(\bs{\omega}), i = 1, 2$ are Gaussian spectral processes with $E[Z_i(\bs{\omega})] = 0$, $V[Z_i(\bs{\omega})] = \sigma_i^2f_i(\bs{\omega})$, and $Cov[Z_i(\bs{\omega}),Z_j(\bs{\omega}')] = 0, i,j = 1,2 $ for $ \bs{\omega}\ne \bs{\omega}' $. The dependence between the two processes is captured via their correlation at each frequency   $Cor[Z_1(\bs{\omega}),Z_2(\bs{\omega})]=\phi(\bs{\omega}) = \phi(-\bs{\omega}) \in (-1,1) $. For example, if the two processes are highly correlated at the low frequency (small $||\bs{\omega}||$), they should exhibit similar large-scale spatial trends; if high correlation exits at the high frequency (large $||\bs{\omega}||$), then the two process should exhibit similar small-scale spatial features. 
	
	Since CMAQ data $X$ are known at all grid points, we consider the conditional model of monitoring data given CMAQ. In the spectral domain, we can write the conditional distribution of the spectral processes for each frequency as follows,
	\begin{equation}
	\begin{aligned}
	Z_2(\bs{\omega}) &= \alpha(\bs{\omega})Z_1(\bs{\omega}) + Z^\ast(\bs{\omega}), 
	\end{aligned}
	\end{equation}
	where $\alpha(\bs{\omega}) = \phi(\bs{\omega})\frac{\sigma_2\sqrt{f_2(\bs{\omega}})}{\sigma_1\sqrt{f_1(\bs{\omega})}}$, $Z^\ast(\bs{\omega})$ is a Gaussian variable with $E[Z^\ast(\bs{\omega})]=0$ and $Var[Z^\ast(\bs{\omega})] \le Var[Z_2(\bs{\omega})]$, independent of $ Z_1(\bs{\omega}) $ and independent over $ \bs{\omega} $. In the spatial domain, the conditional distribution of $Y(\bs{s}) | X(\bs{s}^\prime) \text{ for all } \bs{s}^\prime$ integrates over all frequencies,
	\begin{equation}
	\begin{aligned}
	Y(\bs{s}) | X(\bs{s}^\prime) \text{ for all } \bs{s}^\prime &= \int \exp(i\bs{\omega}^T\bs{s})\alpha(\bs{\omega})Z_1(\bs{\omega})d\bs{\omega}\\
	& + \int \exp(i\bs{\omega}^T\bs{s})Z^\ast(\bs{\omega})d\bs{\omega},
	\end{aligned}
	\end{equation}
	where the first term is the conditional mean of $Y(\bs{s})$ given $ X $, and the second term is a mean-zero Gaussian process that represents the remaining variability not captured $X$. 
	
	\subsection{Multivariate Spectral Downscaler}
	We extend the univariate spectral downscaler to the multivariate setting, such that the cross-species association between CMAQ and station data can be estimated from the data. Let $\bs{X}(\bs{s}) \in \mathcal{R}^J$ and $ \bs{Y}(\bs{s}) \in \mathcal{R}^K$ be vectors of the multiple pollutants from CMAQ and stations, respectively. A joint spectral model for $\bs{X}(\bs{s})$ and $\bs{Y}(\bs{s})$ is,
	
	\begin{equation}
	\begin{aligned}
	\bs{X}(\bs{s}) &= \int \exp(i\bs{\omega}^T\bs{s})\bs{Z_1}(\bs{\omega})d\bs{\omega} \\
	\bs{Y}(\bs{s}) &= \int \exp(i\bs{\omega}^T\bs{s})\bs{Z_2}(\bs{\omega})d\bs{\omega} ,
	\end{aligned}
	\end{equation}
	where $\bs{Z_1}(\bs{\omega}) \in \mathcal{R}^J$ and $\bs{Z_2}(\bs{\omega}) \in \mathcal{R}^K$ are mean-zero Gaussian vector with $J\times J$ and $K\times K$ covariance matrix $\Sigma_{ii}(\bs{\omega})$  for i=1 and 2 respectively, and $J\times K$ covariance matrix $\Sigma_{12}(\bs{\omega})$. 
	
	Similar to the univariate case, we model station data conditional on CMAQ. It can be shown that the conditional model at frequency $\bs{\omega}$ is,	 \begin{equation}
	\begin{aligned}
	\bs{Z}_2(\bs{\omega}) &= A(\bs{\omega})\bs{Z}_1(\bs{\omega}) + \bs{Z}^\ast(\bs{\omega}),
	\end{aligned}
	\end{equation}
	where $ A(\bs{\omega}) = \Sigma_{21}(\bs{\omega}){\Sigma_{11}^{-1}(\bs{\omega})}$ is a $K \times J$ matrix, and  $\bs{Z}^\ast(\bs{\omega}) \in \mathcal{R}^K$ is a K-dimensional mean-zero Gaussian vector with $ \text{Cov}[\bs{Z}^\ast(\bs{\omega})] = \Sigma_{22}(\bs{\omega}) - \Sigma_{21}(\bs{\omega})\Sigma_{11}^{-1}(\bs{\omega})\Sigma_{12}(\bs{\omega})$ and independent of $\bs{Z}_1(\bs{\omega})$. 
	The resulting conditional model in the spatial domain is
	\begin{equation}\label{eq:multispatial}
	\begin{aligned}
	\bs{Y}(\bs{s}) | \bs{X}(\bs{s}^\prime) \text{ for all } \bs{s}^\prime &= \int \exp(i\bs{\omega}^T\bs{s})A(\bs{\omega})\bs{Z}_1(\bs{\omega})d\bs{\omega}\\
	& + \int \exp(i\bs{\omega}^T\bs{s})\bs{Z}^\ast(\bs{\omega})d\bs{\omega},
	\end{aligned}
	\end{equation}
	where the first term is the conditional mean $\bs{\mu(s)} = E[\bs{Y}(\bs{s}) | \bs{X}(\bs{s}^\prime) \text{ for all } \bs{s}^\prime]$ and the second term is a mean-zero stationary multivariate Gaussian process.
	
	Since the regression coefficient matrix $ A(\bs{\omega}) $ between the two spectral processes are unknown at every frequency, extending \cite{Reich2014} to the multivariate setting, we model each element of the $ K\times J$ matrix as a linear combination of $ B $ B-spline basis functions $\mathcal{B}_{1}(\bs{\omega}), \dots, \mathcal{B}_{B}(\bs{\omega})$, 
	\begin{equation}\label{eq:basis}
	A_{kj}(\bs{\omega}) = \sum_{b=1}^{B} \beta_{kjb}\mathcal{B}_{b}(\bs{\omega}).
	\end{equation}
	Substituting \eqref{eq:basis} into \eqref{eq:multispatial}, the conditional mean becomes $ \mu_k(\bs{s})=  \sum_{j=1}^{J} \sum_{b=1}^{B} \tilde{X}_{jb}(\bs{s})\beta_{kjb}$, where $\tilde{X}_{jb}(\bs{s})=\int \mathcal{B}_{b}(\bs{\omega})\exp(i\bs{\omega}^T\bs{s})Z_{1j}(\bs{\omega})d\bs{\omega}$ are the constructed spectral covariates. For data on a grid, the integration is approximated by $\tilde{X}_{jb} = \sum_{l=1}^{M}\mathcal{B}_{b}(\bs{\omega}_l)\exp(i\bs{\omega}_l^T\bs{s})Z_{1j}(\bs{\omega}_l)$, which can be computed efficiently by applying fast Fourier transform (FFT) twice. We first obtain $\hat{Z}_{1j}(\bs{\omega}_l), l=1,\dots,M$ by applying inverse FFT on $X_j$, then weight the signals by multiplying the b$^{th}$ basis function evaluated at $||\bs{\omega}_l||$ and finally applying FFT on the product $\hat{Z}_{1j}(\bs{\omega}_l)\mathcal{B}_{b}(\bs{\omega}_l)$. 
	
	Since CMAQ data are known, all the spectral covariates only need to be computed once prior to model fitting, then the multivariate spectral downscaler can be written as a multivariate spatial regression model. To simplify notation, let $ \bs{Y}_k = \left[ Y_k(\bs{s}_1),\dots,Y_k(\bs{s}_{n_k}) \right] ^T $ collect all station data for the $ k $-th pollutant, and let $ \tilde{\bs{X}}_{jb} = \left[ \tilde{X}_{jb}(\bs{s}_1),\dots,\tilde{X}_{jb}(\bs{s}_{n_k})  \right] ^T$ be the collection of spectral covariates for the $ j $-th pollutant and $ b $-th basis at the grid cell containing the station locations. Then the multivariate spatial regression model for the observations is
	\begin{equation}\label{eqn:model}
	\bs{Y}_k = \beta_{k_0} +  \sum_{j=1}^{J}\sum_{b=1}^{B}\beta_{kjb}\tilde{\bs{X}}_{jb} +\bs{w}_k + \bs{\epsilon}_k, k = 1,\dots,K, 
	\end{equation}
	where $\bs{w} = [\bs{w}_1^T,\dots,\bs{w}_K^T]^T $ are multivariate spatial random effects to capture the remaining dependence (details provided in the next section), and $\bs{\epsilon}_k$ are independent measurement errors corresponding to the k-th pollutant, $ \bs{\epsilon}_k \sim N(0,\tau_k^2\bs{I})$.
	
	\subsection{Multivariate Spatial Model}
	Exploratory analysis of the residuals after accounting for the spectral covariates suggests they have negligible temporal dependence. We model $\bs{w}(\bs{s}) = [w_1(\bs{s}),\dots,w_K(\bs{s})]^T$ as a purely-spatial multivariate process. Here we introduce the process as if there are no missing data at each location, but the covariance matrix for $\bs{w}$ the stacked random effects with missing data is formed explicitly during model fitting. A commonly-used and easy-to-interpret model for multivariate spatial process is the linear model of coregionalization (LMC) proposed by Gelfand et al., (2004), where the process is expressed as a linear combination of independent spatial processes $\bs{w}(\bs{s}) = L\bs{v}(\bs{s})$. Without loss of generality, $L$ is taken as a $ K\times K $ lower triangular matrix and $\bs{v}(\bs{s}) = [v_1(\bs{s}),\dots,v_K(\bs{s})]^T$ with $v_k(\bs{s})\stackrel{ind}{\sim} GP(0,\rho_k(\cdot,\phi_k))$. That is, $\text{Cov}(v_k(\bs{s}), v_{k^\prime}(\bs{s}^\prime))$ is zero for $ k\ne k^\prime $, and $ \rho_k(||\bs{s}- \bs{s}'||, \phi_k)$ otherwise; the spatial dependence of each process is determined by the covariance function $ \rho_k $. For convenience, we use the exponential covariance function, $ \rho_k(||\bs{s}- \bs{s}'||,\phi_k) = \exp({-{\phi_k}||\bs{s}- \bs{s}'||})$, which assumes a decreasing spatial dependence as the Euclidean distance $ ||\bs{s}- \bs{s}'|| $ increases with the rate of decay controlled by the spatial decay parameter $ \phi_k $. 
	
	This model is easy to interpret as the cross dependence and spatial dependence for the multivariate spatial process is explicit. For any location $ \bs{s} $, the variance-covariance matrix for the cross dependence is $ C = LL^T$, i.e., $\text{Cov} \left\lbrace {w}_i(\bs{s}),{w}_j( \bs{s}) \right\rbrace = C_{ij}$. The spatial dependence within the i-th pollutant is $ \text{Cov} \left\lbrace {w}_i( \bs{s}),{w}_i( \bs{s}') \right\rbrace = \sum_{j=1}^{K}L_{ij}^2\rho_{j}( ||\bs{s}- \bs{s}'||, \phi_{j})$, and between the i-th and j-th pollutants is $\text{Cov} \left\lbrace {w}_i( \bs{s}),{w}_j( \bs{s}') \right\rbrace   = \sum_{k=1}^{K} L_{ik}L_{jk}\rho_k(||\bs{s}- \bs{s}'||,\phi_k)$. For a model with missing data, we can denote the observation for a pollutant by a vector and stack all vectors to form a larger vector as shown in Equation \eqref{eqn:model}; then, the elements of covariance matrix for $\bs{w}$ are explicit. 
	Furthermore, this LMC model provides a spectral representation for the remaining variability after accounting for $\bs{X}$. The second term $\bs{Z}^\ast(\bs{\omega})$ in Equation (7) is therefore a mean-zero stationary multivariate Gaussian process with Cov[$\bs{Z}^\ast(\bs{\omega})$]=$ L{\Lambda(\bs{\omega})}L' + D(\bs{\omega})$, where $\Lambda(\bs{\omega})$ is a diagonal matrix with elements $f_k(\bs{\omega})$ which are the spectral densities corresponding to $v_k(\bs{s})$, and $D(\bs{\omega})$ is a diagonal matrix with elements $\tau^2_k$. 
	
	To complete the Bayesian model, we have the following prior specification. Both the additive bias $ \bs{\beta}_0 = (\beta_{10},\dots,\beta_{K0})^T$ and the mean parameters $\bs{\beta} = \left\lbrace \beta_i: i=1,\dots,p\right\rbrace $ with dimension $p = KJB$ have independent normal priors $ N(0, 100^2)$. 
	For the coregionalization matrix $ L $, we have normal priors $N(0,10^2)$ for the off-diagonal entries and lognormal priors $LN(0,10^2)$ for the $K$ diagonal entires. Lastly, we use inverse gamma priors for the nuggets $\tau_k\stackrel{iid}{\sim} \text{invGamma}(\text{shape}=2,\text{scale}=0.1) $ and a uniform prior for the spatial decay parameters $\phi_k = \phi\sim U \left( \frac{3}{0.75\times d},\frac{3}{0.1\times d} \right)$ where $d$ is the maximum distance in the spatial domain; therefore, the prior corresponds to the effective range (defined as the distance at which spatial correlation drops to 0.05) between $ 0.1\times d $ and $ 0.75\times d $. 
	
	\subsection{Computation}\label{sec:computation}
	The Bayesian hierarchical model is fit using a MCMC algorithm that involves Gibbs sampling for updating $\bs{\beta}_{0}$, $\bs{\beta}$, and $\tau_1^2,\dots,\tau_K^2$, and Metropolis-Hastings sampling for updating the common range parameter $\phi$ and entries of $ L $. We fit the model to 3-day batches in parallel, then combine the MCMC samples using the consensus Monte Carlo algorithm \citep{ConsensusMCMC} provided in the R package \textit{parallelMCMCcombine} \citep{parallelMCMC} for inference.
	Parallel MCMC for ``divide and conquer" to leverage parallel computing have been proposed for data with independent subsets \citep{Wang2013ParallelizingMV,Neiswanger, ConsensusMCMC}. It partitions the data into small independent batches and performs MCMC sampling independently to each batch, then combines the samples to give samples from the approximate full-data posterior for inference. Since CMAQ model has captured most temporal variability in the observations, and the observations are collected typically 1-in-3 and 1-in-6 days, it is reasonable to assume that the multivariate spatial residuals are independent in time and thus the 3-day batches are independent. 
	
	Let $\bs{x}^T = \left\lbrace x_1,\dots, x_T \right\rbrace$ be the full data from a probability density function $p(x|\theta)$ with model parameter $\theta$. If the T observations are conditionally independent and partitioned into $M$ non-overlapping batches, $\left\lbrace \bs{x}^{t_1},\dots,\bs{x}^{t_M} \right\rbrace$, then the full-data posterior and the partition subposterior distributions have the following relationship,
	\begin{equation*}
	\begin{aligned}
	p(\theta|\bs{x}^T) \propto p(\bs{x}^T|\theta)p(\theta)  =  \prod_{t=1}^T p(x_t|\theta)p(\theta) = \prod_{m=1}^M p(\bs{x}^{t_m}|\theta)p(\theta)^{1/M} \propto \prod_{m=1}^M p(\theta|\bs{x}^{t_m}).
	\end{aligned}
	\end{equation*} 
	We can obtain MCMC samples $\theta_{m}^{(i)}, i = 1,\dots,I$ from the subposterior distributions $p(\theta|\bs{x}^{t_m})$, $m = 1,\dots,M$, where each MCMC sample is obtained independently. 
	\cite{ConsensusMCMC} proposed to combine the samples using weighted averages to form a set of draws $\theta^{(i)}$  from the full-data posterior by letting $\theta^{(i)} = \left(\sum_{m=1}^M \Sigma_m^{-1}\right)^{-1} \sum_{m=1}^M \Sigma_m^{-1} \theta_{m}^{(i)}$, where $\Sigma_m = Var(\theta|\bs{x}^{t_m})$ is estimated by computing the sample covariance of $\theta_{m}^{(1)},\dots,\theta_{m}^{(I)}$. The full posterior samples are obtained by approximating the subposterior densities with a multivariate normal density function which is optimal when the subposteriors are near Gaussian or when the sample size is large. In our implementation we found that the subposteriors of the regression coefficients satisfy the Gaussian assumption while the spatial parameters tend to be right skewed. However, the inference for regression coefficients and the prediction performance are not affected by the skewness as illustrated by a simulation (see Supplementary Materials S.4).

	\section{Analysis of Speciated $\PM$ Across the US}
	
	\subsection{Exploratory Analysis}
	We begin by exploring the relationship between station and CMAQ data at different spatial scales by estimating their associations at different frequency bands. For each CMAQ pollutant output $X_j, j = 1,\dots,J$ , we obtain $\tilde{X}_{j,[\delta_l,\delta_{l+1})}$ using (\ref{eqn:filterX}) for 8 equal-width bins in the interval $[0,8\pi/5)$ with $\delta_1=0$ and $\delta_{8}=8\pi/5$, this is chosen to cover the range of available frequency with magnitude in $[0,\sqrt{2}\pi]$. Then, we estimate the conditional mean by least squares assuming residual independence, 
	\begin{equation}\label{eqn:mean}
	E[Y_{k}(\bs{s}) | X_{j}(\bs{s}^\prime) \text{ for all } \bs{s}^\prime \text{and } j] = \beta_{k_0} + \sum_{j=1}^{J}\sum_{l=1}^{10}\beta_{kjl}\tilde{X}_{j,[\delta_l,\delta_{l+1})}(\bs{s}), \hspace{0.5cm} k= 1, \dots,K.
	\end{equation}
	All cross-species predictors are included in the regression mean, for example, EC CMAQ output is used to predict total $\PM$ station data, allowing potential complex relationship among pollutants to be estimated. The coefficient estimates from the regression of station data on spectral covariates with cross-species (which have been standardized) are shown in Table 1 in the Supplementary Materials S.3; the coefficients measure the association between station and CMAQ data at multiple spatial scales. Cross-species predictors are significant (highlighted in blue), for instance, CMAQ sulfate at large scale (240km) is a significant predictor for $\PM$ suggesting that regional average sulfate is predictive of local total $\PM$.
	
	We have used spatially constant slopes $\beta_{kjl}$ in the conditional mean, which assumes the relationship between the station data and spectral covariates is fixed across contiguous US. This assumption is verified by estimating Equation \eqref{eqn:mean} separately for 5 US regions: northeast, southeast, midwest, southwest and west (see Supplementary Materials S.3 for region partitions). The differences in the estimated association are mostly insignificant (see Supplementary Materials S.3), therefore in the remaining analysis we combined all regions for mean estimation. 
	
	We also analyze the spatiotemporal as well as cross-species dependence of the residuals to verify the assumptions that are made implicitly in the model and model fitting. These include temporal independence after accounting for spectral covariates, isotropic and stationary assumptions implied by the exponential covariance function, and further more assuming the same spatial decay parameter for all pollutants. Details on data exploratory analysis are shown in Supplementary Materials S.3, while a summary is presented here. To assess time dependence, we compute the autocorrelation function (ACF) for each pollutant at each site. For $\PM$, the proportions of significant lags at lag-1 and -2 are relatively large, however all constituent species lack daily data to estimate dependence at these lags; for lag-3 and beyond, the autocorrelations are mostly insignificant across all sites for all pollutants. Therefore, it is reasonable to assume residual temporal independence and this agrees with the assumption in Rundel et al (2015). To assess spatial dependence of each pollutant, the daily empirical variogram is computed and a variogram assuming an exponential model is fit to the daily empirical vaiograms combined; we see that spatial range is estimated to be over 100 miles for all pollutants, except NO$_3$ has a smaller range of 29 miles. To assess cross-dependence among pollutants, empirical cross-correlogram is plotted for each pair of pollutants, and it shows that the cross-dependence among pollutants within  distance $<$50 miles could be exploited to improve prediction.
	
	
	\subsection{Data Analysis}
	We assess model performance on spatial interpolation and prediction using cross-validation. For this, we randomly split the monitors into five folds. Each fold has roughly 20\% each from monitoring sites measuring $\PM$ only, speciated $\PM$ only, and both. Four different mean structures are compared: the CMAQ covariate of a single pollutant, the CMAQ covariates with cross-species, the spectral covariates of a single pollutant and the spectral covariates with cross-species. The first two models with mean structures that are linear in CMAQ are called linear downscaler (LD), while the latter two are linear in spectral covariates are called spectral downscaler (SD). We also compare the model with and without residual dependence (i.e. $w_k(\bs{s})=0$); hence, eight models are considered. Models 1-4 assume independent errors and are fit by least squares using the \textit{lm} function in \textit{R}, while models 5-8 take into account spatial and cross dependence and are fit using MCMC as described in Section \ref{sec:computation}. The eight models are: 
	
	\begin{align*}
	&\textbf{LD:} & Y_{k}(\bs{s})  &= \beta_{k_0} +  \beta_kX_k(\bs{s}) + \epsilon_k(\bs{s})\\
	&\textbf{LD + Cross:} & Y_{k}(\bs{s})  &= \beta_{k_0} +  \sum_{j=1}^J\beta_{kj}X_{j}(\bs{s}) + \epsilon_k(\bs{s}) \\
	&\textbf{SD:} & Y_{k}(\bs{s})  &= \beta_{k_0} +  \sum_{b=1}^{B}\beta_{kb}\tilde{X}_{kb}(\bs{s}) + \epsilon_k(\bs{s})\\
	&\textbf{SD + Cross:}  & Y_{k}(\bs{s}) &= \beta_{k_0} +  \sum_{j=1}^J\sum_{b=1}^B\beta_{kjb}\tilde{X}_{jb}(\bs{s}) + \epsilon_k(\bs{s})\\
	&\textbf{Spatial LD:}  & Y_{k}(\bs{s})  &= \beta_{k_0} +  \beta_kX_k(\bs{s}) + w_k(\bs{s}) + \epsilon_k(\bs{s})\\
	&\textbf{Spatial LD + Cross:}& Y_{k}(\bs{s})  &= \beta_{k_0} +  \sum_{j=1}^J\beta_{kj}X_{j}(\bs{s}) + w_k(\bs{s}) + \epsilon_k(\bs{s})\\
	&\textbf{Spatial SD:}   & Y_{k}(\bs{s})  &= \beta_{k_0} +  \sum_{b=1}^B\beta_{kb}\tilde{X}_{kb}(\bs{s}) + w_k(\bs{s}) + \epsilon_k(\bs{s})\\
	&\textbf{Spatial SD + Cross:} & Y_{k}(\bs{s})  &= \beta_{k_0} +  \sum_{j=1}^J\sum_{b=1}^B\beta_{kjb}\tilde{X}_{jb}(\bs{s}) + w_k(\bs{s}) + \epsilon_k(\bs{s})\\
	\end{align*}
	
	We fit all eight models separately for each season (Jan-Mar, Apr-Jun, July-Sep and Oct-Dec), allowing the regression coefficients and dependence structure to vary. For each season, the first 78 days of the season is the training period and the last 12 days is the testing period. Two types of cross-validation are performed for comparing model performance:  interpolation/spatial prediction and forecast/temporal prediction. The former is computed over the training period and evaluates how well a model predict at locations without an observation; this is of interest, for example, in epidemiology research to study the effects of long-term exposures to $\PM$ on health outcomes \citep[e.g.]{Adams2015}. The latter is computed over the testing period and evaluates model performance in making short-term forecasts for quantifying air pollution levels at local and national scales.
	
	For interpolation, we predict at the hold-out sites in each fold using the fitted model trained on the other 4 folds. Both hold-out and training data are observed in the training period. We compare the predicted values with the hold-out observations by averaging the root mean squared errors (RMSEs) and correlations across 5 folds. The performance of the spatial models are better than the independent models, due to the borrowing strength from neighboring observations; therefore, Table \ref{tab:interp} shows only the interpolation result of the spatial models. The models are comparable across all species; this indicates that when spatial and cross dependence among species are accounted for, increasing the complexity of the mean structure may not improve interpolation. Among all pollutants, the interpolation for EC has the lowest correlation, this is likely because EC is a marker of traffic emissions which typically have high spatial heterogeneity \citep{pmguide}. 
	
	For temporal prediction, the model is trained using data from each 4 folds in the training period, and the fitted model predicts at all sites for the testing period. In this case, the independent models perform better than the spatial models, and their results are presented in Table \ref{tab:interp}. Including cross-species predictors improves prediction, for instance, {LD + Cross} outperforms {LD} and {SD + Cross} outperforms {SD} for all species except SO$_4$. The SD+Cross model also gives the lowest RMSEs for all species except SO$_4$. This suggests that when spatial and cross-species dependence are not taken in to account, increasing mean complexity improves prediction.

\begin{table}[h]
	\centering
	\caption{\textbf{Root mean squared error ($\mu g/m^3$) and correlation (parenthesis)} between observations and predicted values from multipollutant interpolation and prediction averaged over all seasons of 2011. Interpolation results of the spatial models are presented; they are the spatial linear downscaler (``SpLD") with and without CMAQ cross-species covariates, and the spatial spectral downscaler (``SpSD") with and without cross-species spectral covariates. Temporal prediction results of the independent models are presented; they are the linear downscaler (``LD") with and without CMAQ cross-species covariates, and the spectral downscaler (``SD") with and without cross-species spectral covariates. The results are based on 5-fold cross-validation.}
	\label{tab:interp}
	\begin{tabular}{lcccccc}
		\Hline
		Interpolation&PM25&EC&NO3&NH4&OC&SO4\\
		\hline
		{SpLD}&2.56(0.90)&0.44(0.58)&0.63(0.91)&0.41(0.87)&0.79(0.80)&0.75(0.88)\\
		{SpLD + Cross} &2.56(0.90)&0.45(0.57)&0.66(0.91)&0.40(0.88)&0.79(0.80)&0.73(0.88)\\
		{SpSD}&2.57(0.90)&0.44(0.59)&0.63(0.91)&0.41(0.87)&0.79(0.80)&0.72(0.89)\\
		{SpSD +
			Cross}&2.56(0.90)&0.45(0.56)&0.74(0.88)&0.58(0.79)&0.83(0.77)&0.83(0.85)\\
		\hline
				Prediction&PM25&EC&NO3&NH4&OC&SO4\\
		 \hline
		{LD}        &5.11(0.40)&0.41(0.53)&2.18(0.57)&0.84(0.58)&1.26(0.46)&0.95(0.69) \\
		{LD + Cross}&4.89(0.46)&0.41(0.56)&1.87(0.73)&0.80(0.60)&1.25(0.47)&0.98(0.66) \\
		{SD}        &5.06(0.41)&0.42(0.53)&2.14(0.60)&0.87(0.50)&1.28(0.45)&0.94(0.70) \\
		{SD + Cross}&4.80(0.49)&0.40(0.58)&1.81(0.76)&0.79(0.63)&1.19(0.55)&0.94(0.69)\\
		\hline
	\end{tabular}
\end{table}

\begin{figure}
		\centering
		\subfloat[LD]{\includegraphics[page=1,width=0.5\textwidth]{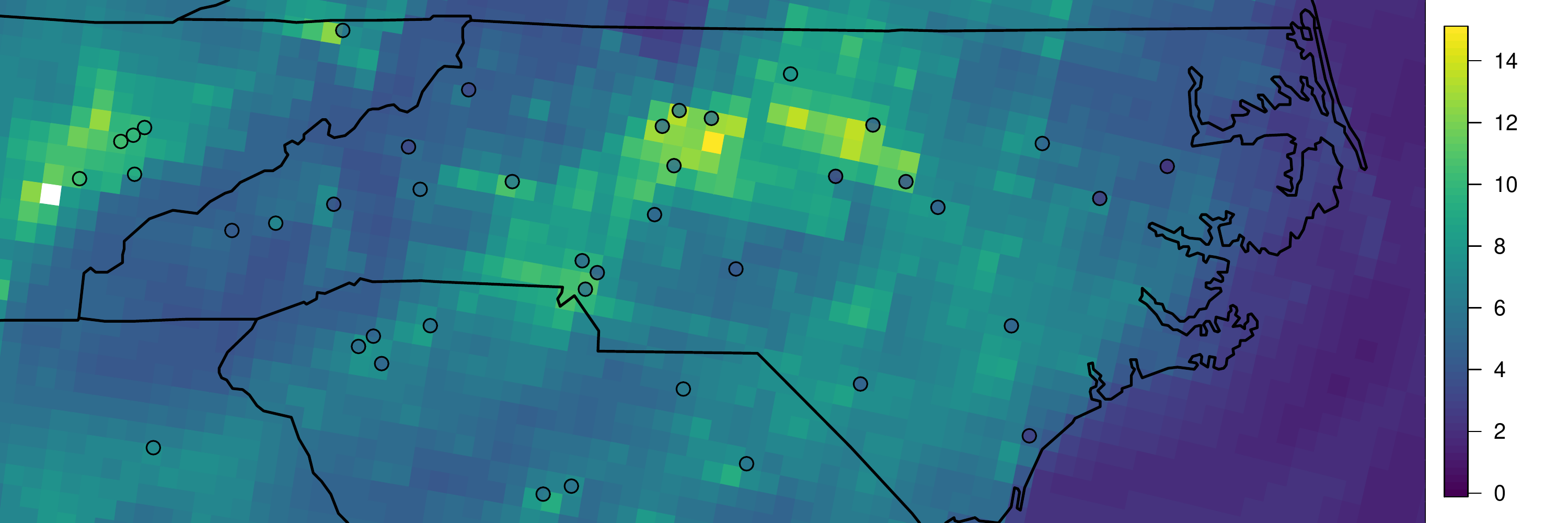}}
		\subfloat[LD + Cross]{\includegraphics[page=2,width=0.5\textwidth]{mappredday264exp_PM25_zoom.pdf}}
		
		\subfloat[SD]{\includegraphics[page=3,width=0.5\textwidth]{mappredday264exp_PM25_zoom.pdf}}
		\subfloat[SD + Cross]{\includegraphics[page=4,width=0.5\textwidth]{mappredday264exp_PM25_zoom.pdf}}
		
		\subfloat[$\PM$ CMAQ]{\includegraphics[page=5,width=0.5\textwidth]{mappredday264exp_PM25_zoom.pdf}}
		\caption{\textbf{Predicted $\PM$ ($\mu g/m^3$) for North Carolina} on Sept 21, 2011 from the linear downscaler (``LD") with and without CMAQ cross-species covariates, and the spectral downscaler (``SD") with and without cross-species spectral covariates. CMAQ $\PM$ output and the station observation (colored circles) are plotted for comparison. The color scales for figures (a-b), (c-d) and (e) are different in order to show the pollution pattern.}\label{fig:mappred}
	\end{figure}

	Figure \ref{fig:mappred} shows the $\PM$ predictions for North Carolina (NC) on September 21, 2011 from the independent models, as well as CMAQ $\PM$ and station data for comparison. The maps show a similar overall pattern, however the LD, LC + Cross and SD models overestimate $\PM$ for central NC while SD + Cross seems to correct the overestimation. 
	The predicted map from SD + Cross model is smoother comparing to LD and LD + Cross models, suggesting that only the large spatial scale features from CMAQ output are used in prediction. 
	
\begin{figure}
		\centering
		\subfloat[January-March]{\includegraphics[page=1,width=0.9\textwidth]{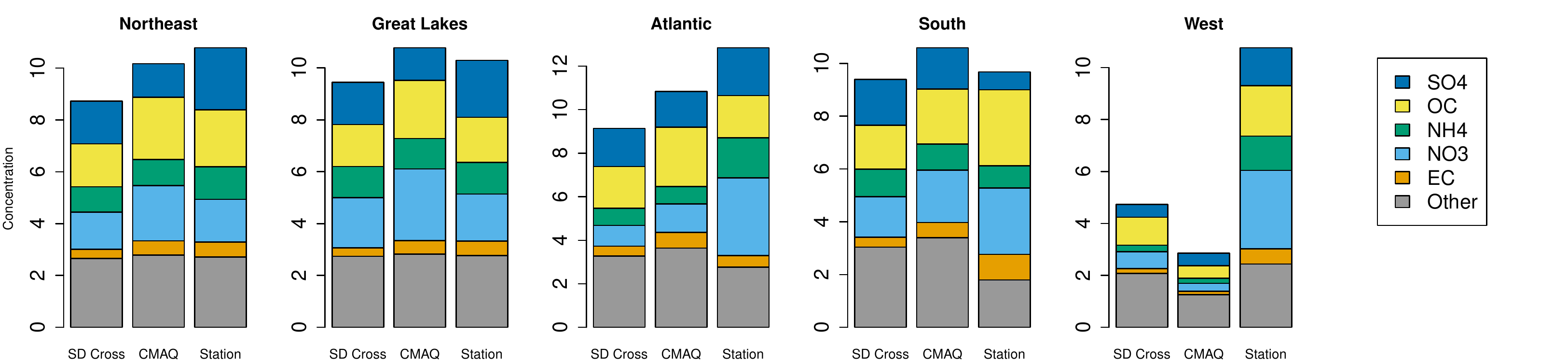}}
		
		\subfloat[April-June]{\includegraphics[page=2,width=0.9\textwidth]{barplot_region_season.pdf}}
		
		\subfloat[July-September]{\includegraphics[page=3,width=0.9\textwidth]{barplot_region_season.pdf}}
		
		\subfloat[October-December]{\includegraphics[page=4,width=0.9\textwidth]{barplot_region_season.pdf}}
		\caption{\textbf{Multipollutant concentrations ($\mu g/m^3$) by US region and season.} Results from spectral downscaler with cross-species predictors (``SD Cross") and CMAQ outputs are averaged within the region and season. Station data are averaged across available monitoring sites and days within the region and season. The ``Other" pollutant is computed by taking the difference of  total $\PM$ and the sum of 5 constituent species. The five regions are grouped by states with similar $\PM$ sources derived from principle component analysis according to the EPA \citep{CMAQeval}. Northeast (Maine, New Hampshire, Vermont, Massachusetts,New York, New Jersey, Maryland, Delaware, Connecticut, Rhode Island, Pennsylvania, District of Columbia, Virginia and West Virginia), Great Lakes (Ohio, Michigan, Indiana, Illinois and Wisconsin), Atlantic (North Carolina, South Carolina, Georgia and Florida), south (Kentucky, Tennessee, Mississippi, Alabama, Louisiana, Missouri, Oklahoma and Arkansas) and west (California, Oregon, Washington, Arizona, Nevada and New Mexico). }\label{fig:barplot}
	\end{figure}

	To gain better insight on the spectral downscaler performance, Figure \ref{fig:barplot} shows the multipollutant concentrations by US region and season for 2011. Daily multipollutant concentration maps for the entire US are interpolated from spatial SD + Cross for the training period and predicted from the SD + Cross for the testing period. Then, both map prediction and the CMAQ outputs are averaged within the region and season, while station data are averaged across available monitoring sites and days within the region and season for comparison. CMAQ predicted lower concentrations for the west for the entire year, and for all region for the months of Apr-Jun (AMJ) and Jul-Sep (JAS) for all pollutants, while the downscaler also underpredicts but less than CMAQ. The overall lower concentration when compared to station data is most likely due to the sampling bias of the monitoring system, as there are more sites located in urban areas with high air pollution due to car emission, coal combustion and industry activities, etc. A comparison of the concentration differences between CMAQ and station data across region and season reveals that CMAQ tends to underpredict in the west region, while the downscaler alleviates the underprediction issue. 
	
\begin{figure}
		\centering
		\subfloat{\includegraphics[page=1,width=0.4\textwidth,trim={0cm 0cm 0cm 1cm},clip]{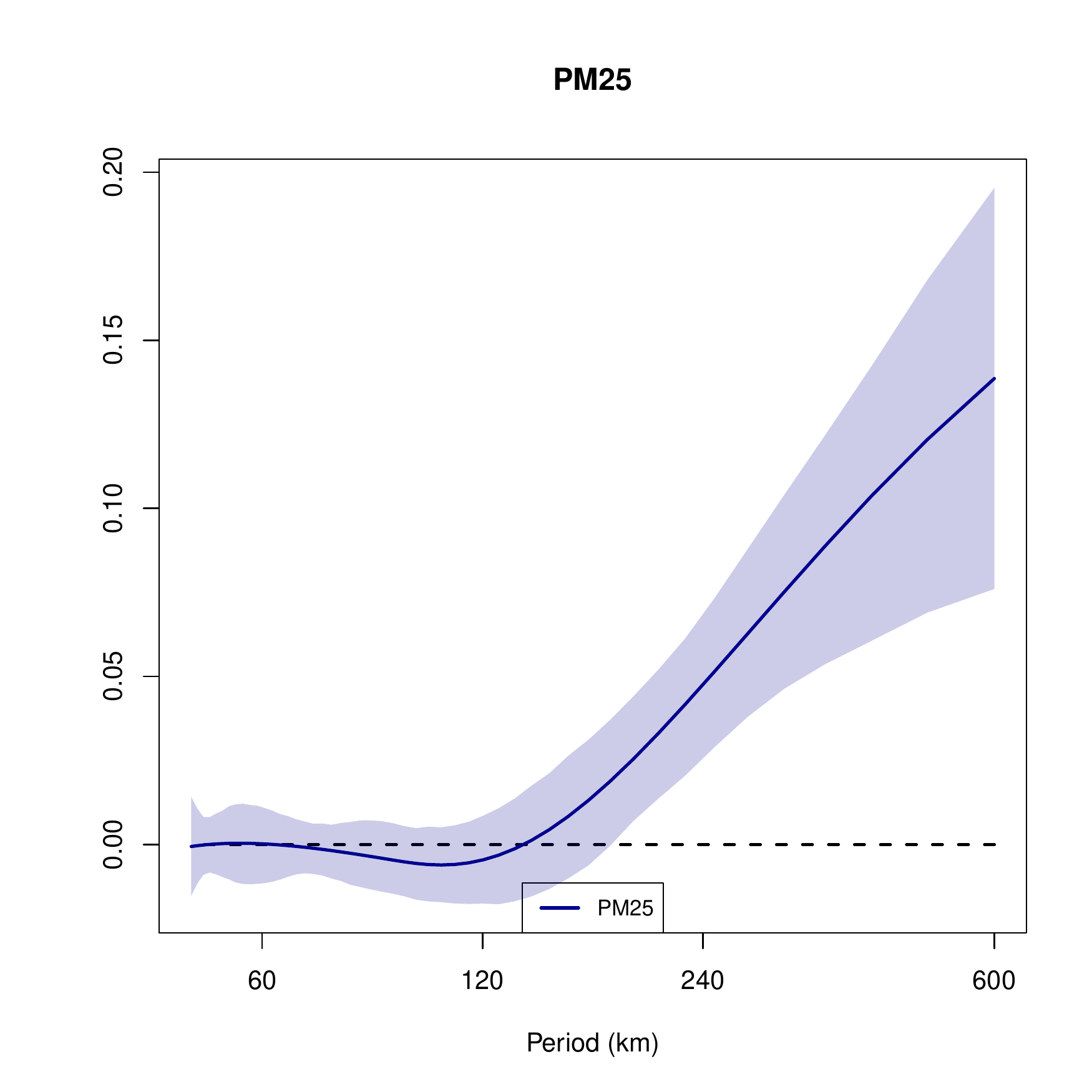}}
		\subfloat{\includegraphics[page=2,width=0.4\textwidth,trim={0cm 0cm 0cm 1cm},clip]{coefJAS_bonf_nbasis5_P31_fold2.pdf}}
		
		\subfloat{\includegraphics[page=3,width=0.4\textwidth,trim={0cm 0cm 0cm 1cm},clip]{coefJAS_bonf_nbasis5_P31_fold2.pdf}}
		\subfloat{\includegraphics[page=4,width=0.4\textwidth,trim={0cm 0cm 0cm 1cm},clip]{coefJAS_bonf_nbasis5_P31_fold2.pdf}}
		
		\subfloat{\includegraphics[page=5,width=0.4\textwidth,trim={0cm 0cm 0cm 1cm},clip]{coefJAS_bonf_nbasis5_P31_fold2.pdf}}
		\subfloat{\includegraphics[page=6,width=0.4\textwidth,trim={0cm 0cm 0cm 1cm},clip]{coefJAS_bonf_nbasis5_P31_fold2.pdf}}
		\caption{\textbf{Posterior mean (solid) and 95\% interval (shade) of estimated association} plotted by period. The association function is declared significant if the coefficients are significant at a 5\% significant level with Bonferroni correction.}\label{fig:coef}
	\end{figure}

	Further evaluation of the CMAQ model at different spatial scales can be assessed. Figure \ref{fig:coef} shows the coherence, $A_{kj}(\bs{\omega})$, between station data and CMAQ estimated from the Spatial SD + Cross model fitted to Jul-Sep, 2011. The estimated association is a function of the frequency $\bs{\omega}$ and is converted to spatial scale by using 12km$\times 2\pi/||\bs{\omega}||$, where 12km is the CMAQ output grid size. All observed pollutants are significantly related to their corresponding CMAQ outputs at spatial scale greater than 120km, except NH$_4$. However, for most pollutants local variation in CMAQ (period less then 120km) is not associated with local variation in the station data.  That is, the model is filtering the CMAQ output to remove high-frequency terms. Some observed pollutants are significantly related to cross-species CMAQ outputs, for example, EC is found significantly associated with NO$_3$ and NH$_4$. It may be that these species share common sources that affect air quality at this spatial range, such as emissions from the combustion of various fuel \citep{pmguide}.
	
	\section{Discussion}
	The proposed spectral downscaler exploits the complex relationship between the station and proxy data. As shown in our application, explicitly modeling the association between the two data sources at different spatial scales improves temporal prediction performance for total and speciated $\PM$, while joint modeling the multiple pollutants with multivariate spatial dependence improves spatial prediction. The associations estimated at different spatial scales provide a valuable alternative tool for model evaluation, and this is used to evaluate daily CMAQ multipollutant output for the first time.
	
	We have implicitly assumed stationarity in modeling the station observations. Nonstationarity can be introduced by allowing the mean parameters $\bs{\beta}$ to vary spatially. However this is not implemented in our application, since the mean parameters estimated from least squares are mostly not significantly different by region (see Supplementary Materials S.3). We have also assumed the multivariate spatial random effect $\bs{w}$ is isotropic and stationary, since the mean structure in our model allows for flexible relationships between the response and numerical model, which possibly captures the underlying process that manifest spatial heterogeneity. One way to introduce nonstationarity is to allow the cross dependence matrix to be spatially varying by letting $L(\bs{s})$ vary across location; however, considering the sparsity of the station data, we have chosen a parsimonious model that seems adequate for our application.
	
	In our model fitting, we have used parallel MCMC for divide and conquer large data set. Since standard R packages for fitting multivariate spatial process (without replicates), such as \textit{spBayes} (Finley et al, 2015), and for combining subposterior MCMC samples, such as \textit{parallelMCMCcombine} \citep{parallelMCMC}, are available in \textit{R}. It is convenient for practitioners to apply this model to their applications. The only additional programming is to construct the spectral covariates which only need to be computed once. The data likelihood can be parallelized, one could develop their own code to fit the model to the entire data set without partitioning. R Code to implement the spectral downscaler and Supplementary Materials are available on Github \url{https://github.com/yawenguan/multires}
	
\backmatter	
\section*{Acknowledgements}
The authors would like to thank Dr. Joe Guinness for spectral analysis expertise. This work was supported by the National Institutes of Health (ES027892) and the National Science Foundation (DMS-1638521). The content of this publication is solely the responsibility of the authors and does not necessarily represent the official views of the NIH or the NSF.
\linespread{1}
\bibliographystyle{biom} 
\bibliography{ref}
\end{document}